# Magnetic nanoparticle traveling in external magnetic field

N. A. Usov[1] and B. Ya. Liubimov[1]

[1]Institute of Terrestrial Magnetism, Ionosphere and Radio Wave Propagation, Russian Academy of Sciences, (IZMIRAN), 142190, Troitsk, Moscow, Russia

A set of equations describing the motion of a free magnetic nanoparticle in an external magnetic field in a vacuum, or in a medium with negligibly small friction forces is postulated. The conservation of the total particle momentum, i.e. the sum of the mechanical and the total spin momentum of the nanoparticle is taken into account explicitly. It is shown that for the motion of a nanoparticle in uniform magnetic field there are three different modes of precession of the unit magnetization vector and the director that is parallel the particle easy anisotropy axis. These modes differ significantly in the precession frequency. For the high-frequency mode the director points approximately along the external magnetic field, whereas the frequency and the characteristic relaxation time of the precession of the unit magnetization vector are close to the corresponding values for conventional ferromagnetic resonance. On the other hand, for the low-frequency modes the unit magnetization vector and the director are nearly parallel and rotate in unison around the external magnetic field. The characteristic relaxation time for the low-frequency modes is remarkably long. This means that in a rare assembly of magnetic nanoparticles there is a possibility of additional resonant absorption of the energy of alternating magnetic field at a frequency that is much smaller compared to conventional ferromagnetic resonance frequency. The scattering of a beam of magnetic nanoparticles in a vacuum in a non-uniform external magnetic field is also considered taking into account the precession of the unit magnetization vector and director.

## I. Introduction

Magnetic nanoparticle is a unique physical object since both classical and purely quantum degrees of freedom have to be taken into account to describe its behavior in external magnetic field. The particle angular velocity $\omega$ and tensor of inertia represent the classical degrees of freedom of a nanoparticle. The tensor of inertia is reduced [1] to a scalar $I$ in the simplest case of a spherical nanoparticle. The quantum degrees of freedom are described [2,3] by a macrospin $S$. The latter in the quasi-classical approximation is defined as the ratio of the particle total magnetic moment to the gyromagnetic ratio $\gamma$, $S = - M_sV\alpha/\gamma$, where $M_s$ is the saturation magnetization, $V$ is the particle volume, and $\alpha$ is the unit magnetization vector. According to the general quantum mechanical principle [4], the total momentum of the particle $J$, which is the sum of the mechanical angular momentum, $L = I\omega$, and the total spin momentum $S$, is conserved for an isolated nanoparticle

$$\vec{J} = \vec{L} + \vec{S} = I\vec{\omega} - M_sV\vec{\alpha}/\gamma = const. \qquad (1)$$

It has been shown [5,6] for the first time that the conservation of the total momentum $J$ leads to important consequences for the quantum tunneling of the magnetic moment of an isolated magnetic nanoparticle. However, the probability of quantum tunneling of the magnetic moment is nonzero [7] only for particles of very small diameters, $D \leq 2$-$4$ nm. At the same time, the conservation of the total momentum $J$ may determine the behavior of a magnetic nanoparticle in a much larger range of sizes.

Consider, for example, a spherical $Fe_3O_4$ nanoparticle having the saturation magnetization $M_s = 480$ emu/cm$^3$ and the density $\rho = 5$ g/cm$^3$, respectively [8]. The single-domain diameter of such nanoparticle has been estimated [9] to be $D_c = 64$ nm. It is easy to see that for a single-domain $Fe_3O_4$ particle with a diameter $D = 50$ nm the total spin momentum is given by $S = 1.8 \times 10^{-21}$ erg×s. This value of the spin momentum is relatively large, as the particle mechanical angular momentum, $L = I\omega = \rho VD^2\omega/10$, is compared with the spin momentum only for a sufficiently large value of the particle angular velocity, $\omega = 2 \times 10^6$ 1/s. This unique physical situation realizes mostly for magnetic nanoparticles, as for particles of macroscopic sizes the mechanical angular momentum is overwhelmingly large even at small angular velocities. This is a consequence of the fact that the particle moment of inertia increases proportionally to $D^5$. Of course, the angular velocity of the nanoparticle is generally low [10] in the presence of viscous friction in the environment, for example, in a viscous liquid. In this case one has $L \ll S$, and Eq. (1) does not hold. However, Eq. (1) is crucial in the study of motion of magnetic nanoparticle in an external magnetic field in a vacuum, or in an environment with negligible small friction forces.

In the present paper we postulate the equations that determine the motion of a free single-domain nanoparticle in external magnetic field. It is shown that in a homogeneous magnetic field there are three different modes of precession of the unit magnetization vector $\alpha$ and the unit vector $n$, that shows the direction of the particle easy anisotropy axis. These modes differ significantly in the precession frequency. The frequency of the highest mode $\nu_1$ is close to that of a conventional ferromagnetic resonance, the direction of the vector $n$ being almost parallel to the external magnetic field. Remarkably, however, that for a free nanoparticle there are two low-frequency modes with frequencies $\nu_2$, and $\nu_3 \approx - \nu_2$. For these modes the vectors $\alpha$ and $n$ are nearly parallel and rotate in unison around the direction of the external magnetic field. The low-frequency modes have relatively long relaxation times. Consequently, in a rare assembly of magnetic nanoparticles in a vacuum there exists resonance absorption of the alternating magnetic field at a frequency $\nu_2$, significantly lower than the ordinary ferromagnetic resonance frequency.

In this paper we consider also the scattering of a beam of magnetic nanoparticles in a non-uniform external magnetic field, taking into account the complex motion of the vectors $\alpha$, $n$ and $\omega$. In particular, we show that in the case of scattering of magnetic nanoparticles on a two dimensional magnetic "lens" the trajectory of the magnetic nanoparticle practically coincides with that of passive magnetic moment, whose direction at each point of the trajectory is parallel to the external magnetic field at the same point. This enables one to determine the ratio $M_s/\rho$ by means of the measurement of the scattering angle of a nanoparticle at given initial conditions. This scattering experiment could be made similar to the well-known experiments [11-13] on the scattering of magnetic nanoclusters in external magnetic field.

## II. Basic equations

Consider a spherical uniaxial ferromagnetic nanoparticle that can freely rotate in space in the absence of viscous friction in the environment. In the presence of an external uniform magnetic field $\vec{H}_0$ a total energy of the nanoparticle is given by

$$W = \frac{1}{2} I \omega^2 - K_1 V (\vec{\alpha} \vec{n})^2 - M_s V (\vec{\alpha} \vec{H}_0), \qquad (2)$$

where $K_1$ is the anisotropy constant of the magnetic nanoparticle and other values have been defined above. In a homogeneous external magnetic field the center of mass of the nanoparticle is at rest or moves freely, and this movement presents no interest. On the other hand, the time dependence of the vectors $\vec{\alpha}$, $\vec{n}$ and $\vec{\omega}$ is described by the following equations

$$\frac{d\vec{n}}{dt} = [\vec{\omega}, \vec{n}]; \qquad (3)$$

$$I \frac{d\vec{\omega}}{dt} = \vec{N} - M_s V \kappa \left[ \vec{\alpha}, \left[ \vec{\alpha}, \vec{H}_{ef} - \frac{\vec{\omega}}{\gamma} \right] \right]; \qquad (4)$$

$$\frac{d\vec{\alpha}}{dt} = -\gamma [\vec{\alpha}, \vec{H}_{ef}] - \gamma \kappa \left[ \vec{\alpha}, \left[ \vec{\alpha}, \vec{H}_{ef} - \frac{\vec{\omega}}{\gamma} \right] \right]. \qquad (5)$$

Eq. (3) is usual kinematic relation [1] for a rotating rigid body. Eq. (4) is the equation of motion [1] for the mechanical angular momentum, where $\vec{N}$ is a torque acting on the particle, and Eq. (5) is the Landau - Lifshitz equation [2] for the unit magnetization vector. In Eqs. (4), (5) the effective magnetic field is given by [2]

$$\vec{H}_{ef} = -\frac{\partial W}{V M_s \partial \vec{\alpha}} = \vec{H}_0 + H_k (\vec{\alpha} \vec{n}) \vec{n}, \qquad (6a)$$

where $H_k = 2K_1/M_s$ is the particle anisotropy field. Similarly, the torque in Eq. (4) can be calculated [10] as

$$\vec{N} = \left[ \frac{\partial W}{\partial \vec{n}}, \vec{n} \right] = -2K_1 V (\vec{\alpha} \vec{n}) [\vec{\alpha}, \vec{n}]. \qquad (6b)$$

Note that the set of Eqs. (3) - (6) is consistent in the sense that it leads to the correct equation of motion for the total momentum of the particle

$$\frac{d\vec{J}}{dt} = M_s V [\vec{\alpha}, \vec{H}_0]. \qquad (7)$$

Therefore, the torque applied to the magnetic moment of the nanoparticle, $M_s V \vec{\alpha}$, leads to a change in the particle total momentum, $\vec{J} = \vec{L} + \vec{S}$.

In Eqs. (4) and (5) the terms proportional to the damping constant $\kappa$ are introduced to describe the phenomenological magnetic damping for the motion of the unit magnetization vector.

The structure of the dissipative terms is chosen based on the following arguments. In the absence of damping, $\kappa = 0$, the change of the total particle energy is given by

$$\frac{dW}{dt} = I\vec{\omega}\frac{d\vec{\omega}}{dt} - 2K_1 V(\vec{\alpha}\vec{n})\left(\frac{d\vec{\alpha}}{dt}\vec{n} + \vec{\alpha}\frac{d\vec{n}}{dt}\right) - M_s V\left(\frac{d\vec{\alpha}}{dt}\vec{H}_0 + \vec{\alpha}\frac{d\vec{H}_0}{dt}\right) =$$

$$M_s V\left(\vec{\alpha}\frac{d\vec{H}_0}{dt}\right). \quad (8)$$

Therefore, the change in the total particle energy is determined by the work done by the external magnetic field $H_0$. Evidently, the total energy is conserved, $dW/dt = 0$, if $H_0$ = const. Eq. (8) can be easily proved using the equations of motion (3) - (5) with $\kappa = 0$.

Let us assume now that the effect of dissipation on the motion of the unit magnetization vector $\alpha$ can be described by introducing an additional term

$$\frac{d\vec{\alpha}}{dt} = -\gamma[\vec{\alpha}, \vec{H}_{ef}] + \gamma[\vec{\alpha}, \vec{\eta}], \quad (9)$$

where $\eta$ is an auxiliary vector proportional to the damping parameter $\kappa$. The structure of the new term follows from the condition $|\vec{\alpha}| = 1$. However, we must keep precise Eq. (7) for the total momentum of the nanoparticle $J$. Therefore, the equation for the vector $L$ must be augmented by a similar dissipation term

$$I\frac{d\vec{\omega}}{dt} = \vec{N} + M_s V[\vec{\alpha}, \vec{\eta}]. \quad (10)$$

Using Eqs. (3), (9) and (10), one can calculate the change in the total energy of the particle at $H_0$ = const as

$$\frac{dW}{dt} = -\gamma M_s V \vec{\eta}\left[\vec{H}_{ef} - \frac{\vec{\omega}}{\gamma}, \vec{\alpha}\right]. \quad (11)$$

As a result of the dissipation the total energy of the particle, Eq. (2), decreases and transforms into thermal energy. Therefore, the right-hand side of the expression (11) must be negative definite. This condition can be satisfied by setting

$$\vec{\eta} = \kappa\left[\vec{H}_{ef} - \frac{\vec{\omega}}{\gamma}, \vec{\alpha}\right]. \quad (12)$$

Then, Eqs. (9) and (10) lead to the relations (5) and (4), respectively.

Eqs. (3) - (6) can be solved numerically using the procedure described in the appendix to Ref. 10. However, to clarify the nature of the resulting complex motion of the vectors $\alpha$, $n$ and $\omega$ in the next section we present an approximate analytical solution of the linearized equations of motion for these vectors.

### III. Particle precession

To obtain approximate analytical description of the motion of the vectors $\alpha$, $n$ and $\omega$ under the influence of external uniform magnetic field it is convenient to rewrite the set of Eqs. (3) - (6) in the following equivalent form

$$\frac{d\vec{n}}{dt} = [\vec{\omega}, \vec{n}]; \quad \frac{d}{dt}(\vec{\omega} - \nu_\alpha \vec{\alpha}) = \nu_\xi^2 [\vec{\alpha}, \vec{k}]; \quad (13)$$

$$\frac{d\vec{\alpha}}{dt} = -\nu_k (\vec{\alpha}\vec{n})[\vec{\alpha}, \vec{n}] - \nu_H [\vec{\alpha}, \vec{k}] - \kappa[\vec{\alpha}, [\vec{\alpha}, \nu_H \vec{k} + \nu_k (\vec{\alpha}\vec{n})\vec{n} - \vec{\omega}]].$$

Here we use Eq. (7) instead of Eq. (4). Besides, the external magnetic field is directed along the z-axis of the Cartesian coordinates, so that $\boldsymbol{H}_0 = H_0\boldsymbol{k}$, where $\boldsymbol{k}$ is the unit vector along the z axis. In Eq. (13) the following notations for the characteristic frequencies of the problem are introduced: $\nu_k = \gamma H_k$, $\nu_H = \gamma H_0$, $\nu_\xi^2 = M_s V H_0/I$, and $\nu_\alpha = M_s V/I\gamma$.

For numerical simulation of Eqs. (13) we use the typical parameters of a soft type magnetic nanoparticle: $K_1 = 10^5$ erg/cm$^3$, $M_s = 500$ emu/cm$^3$, $D = 5\times10^{-6}$ cm, and $\rho = 5$ g/cm$^3$. The external magnetic field is given by $H_0 = 10$ Oe. Then, the characteristic frequencies defined above are given by $\nu_k = 7.04\times10^9$ 1/s, $\nu_H = 1.76\times10^8$ 1/s, $\nu_\xi = 2.0\times10^7$ 1/s, and $\nu_\alpha = 2.27\times10^6$ 1/s, respectively. Here we use the value [14] $\gamma = 1.76\times10^7$ 1/(s×Oe) for the gyromagnetic ratio. Thus, the frequency $\nu_k$ is the largest one in the given problem. It is close to the ferromagnetic resonance frequency [14] of motionless nanoparticle, $\nu_k + \nu_H$, oriented along the external magnetic field. As we shall see further, the frequency $\nu_\xi$ is also significant, even though it is two orders of magnitude less than $\nu_k$. Note that the ratios stated among the characteristic frequencies hold for a sufficiently wide range of the particle parameters.

The numerical simulation of Eqs. (13) shows that for arbitrary initial conditions for the vectors $\boldsymbol{\alpha}$, $\boldsymbol{n}$ and $\boldsymbol{\omega}$ the orientations of the unit magnetization vector and director quickly set close to the external magnetic field direction even in a relatively weak external magnetic field. The subsequent evolution resembles a precession of the vectors $\boldsymbol{\alpha}$ and $\boldsymbol{n}$ around the vector $\boldsymbol{k}$. As a result, the z-components of the vectors $\boldsymbol{\alpha}$ and $\boldsymbol{n}$ turn out to be close to unity. Therefore, a linearization of Eqs. (13) can be carried out putting in Eqs. (13) $\alpha_z = n_z = 1$

$$\frac{dn_x}{dt} = \omega_y - \omega_z n_y; \qquad \frac{dn_y}{dt} = \omega_z n_x - \omega_x;$$

$$\frac{d}{dt}(\omega_x - \nu_\alpha \alpha_x) = \alpha_y \nu_\xi^2; \quad \frac{d}{dt}(\omega_y - \nu_\alpha \alpha_y) = -\alpha_x \nu_\xi^2; \quad \frac{d}{dt}(\omega_z - \nu_\alpha) = 0; \quad (14)$$

$$\frac{d\alpha_x}{dt} = \nu_k n_y - (\nu_k + \nu_H)\alpha_y - \kappa\left[(\nu_k + \nu_H - \omega_z)\alpha_x - \nu_k n_x + \omega_x\right];$$

$$\frac{d\alpha_y}{dt} = (\nu_k + \nu_H)\alpha_x - \nu_k n_x - \kappa\left[(\nu_k + \nu_H - \omega_z)\alpha_y - \nu_k n_y + \omega_y\right].$$

From the conservation of the z-component of the reduced total particle momentum, $J_z/I = j_z = $ const, one obtains the relation $\omega_z = \nu_\alpha + j_z = $ const. Then it is easy to see that neglecting the magnetic damping, i.e. in the limit $\kappa \to 0$, the linearized system of Eqs. (14) has the following general solution

$$\alpha_x = A\cos(\nu t + \varphi); \qquad \alpha_y = A\sin(\nu t + \varphi);$$
$$n_x = B\cos(\nu t + \varphi); \qquad n_y = B\sin(\nu t + \varphi); \qquad (15)$$
$$\omega_x = C\cos(\nu t + \varphi); \qquad \omega_y = C\sin(\nu t + \varphi),$$

where the amplitudes $A$, $B$ and $C$ are related by the equations

$$B = \frac{\nu_k + \nu_H - \nu}{\nu_k}A; \qquad C = \left(\nu_\alpha - \frac{\nu_\xi^2}{\nu}\right)A, \qquad (16a)$$

and the eigen-frequency $\nu$ satisfies the dispersion equation

$$\nu^3 - (\omega_z + \nu_k + \nu_H)\nu^2 + [\omega_z(\nu_k + \nu_H) - \nu_k\nu_\alpha]\nu + \nu_k\nu_\xi^2 = 0. \qquad (16b)$$

It can be shown that Eq. (16b) has three real solutions, so that there are three independent precession modes of the vectors $\boldsymbol{\alpha}$, $\boldsymbol{n}$ and $\boldsymbol{\omega}$. Without loss of generality one can assume that the angular frequency of the particle rotation around the $z$ axis is small compared to other characteristic frequencies, i.e. $\omega_z \ll \nu_\alpha \ll \nu_H \ll \nu_k$. Since the frequency $\omega_z$ is determined by the initial conditions, to simplify the calculations we set further $\omega_z = 0$. Then, one obtains from Eq. (16b) the approximate solutions

$$\nu_1 \approx \nu_k + \nu_H;$$

$$\nu_2 \approx \frac{\nu_\xi}{\sqrt{1+\nu_H/\nu_k}} - \frac{\nu_\alpha}{2(1+\nu_H/\nu_k)} + \frac{\nu_k \nu_\xi^2}{2(\nu_k+\nu_H)^2};  \qquad (17)$$

$$\nu_3 \approx -\frac{\nu_\xi}{\sqrt{1+\nu_H/\nu_k}} - \frac{\nu_\alpha}{2(1+\nu_H/\nu_k)} + \frac{\nu_k \nu_\xi^2}{2(\nu_k+\nu_H)^2}.$$

As we mentioned above, the highest eigen-frequency $\nu_1$ is close to the precession frequency of the magnetic moment of a motionless nanoparticle oriented along the external magnetic field [14]. For this mode the transverse components of the vector $\boldsymbol{n}$ are small, because according to Eq. (16a) the amplitude $B_1 \approx 0$. On the other hand, the modes $\nu_2$ and $\nu_3 \approx -\nu_2$ describe the low-frequency precession of the particle magnetic moment. It is easy to see that for the low frequency modes the vectors $\boldsymbol{\alpha}$ and $\boldsymbol{n}$ move in unison and almost parallel to each other, since the ratio $(\nu_k + \nu_H - \nu_2)/\nu_k \approx 1$, and $B_2 \approx A_2$.

To determine the relaxation times of the oscillation modes (17) it is necessary to take into account the terms proportional to the damping constant $\kappa$ in the linearized set of Eqs. (14). Assuming the damping constant small, it is easy to see that the solution of Eqs. (14) has the form

$$\alpha_x = A\exp(-\beta t)\cos(\nu t + \varphi); \qquad \alpha_y = A\exp(-\beta t)\sin(\nu t + \varphi). \qquad (18)$$

Similar formulas hold for the transverse components of the vectors $\boldsymbol{n}$ and $\boldsymbol{\omega}$. In Eq. (18) the parameter $\beta$ is assumed to be proportional to the damping constant $\kappa$, whereas the frequency $\nu$ still satisfies Eq. (16b). Then, in the lowest order with respect to $\kappa$ one obtains from Eqs. (14) the relation

$$[3\nu - 2(\nu_k + \nu_H)]\beta = \kappa(\nu^2 + \nu_\alpha \nu - \nu_\xi^2). \qquad (19)$$

For the high frequency mode Eq. (19) gives the well-known [14] equation $\beta_1 \approx \kappa(\nu_k + \nu_H)$. On the other hand, the decay rate of the low frequency mode $\nu_2$ is given by

$$\beta_2 = \kappa\frac{(\nu_\xi^2 - \nu_2^2 - \nu_\alpha \nu_2)}{2(\nu_k + \nu_H)} \approx \kappa\frac{\nu_\xi^2 \nu_H}{2(\nu_k + \nu_H)^2}. \qquad (20)$$

The same expression holds also for the decay rate of the mode $\nu_3$. Note that the values $\beta_2$, $\beta_3$ turn out to be small, so that the corresponding relaxation times of these modes, $\tau = 1/\beta$, are relatively large. Indeed, assuming $\nu_k = 7.04\times10^9$ 1/s, $\nu_H = 1.76\times10^8$ 1/s and $\nu_\xi = 2.0\times10^7$ 1/s, one obtains from Eq. (20) $\tau_2 = 1.4\times10^{-3}/\kappa$ s, while the relaxation time of the high-frequency mode at the same characteristic frequencies equals to $\tau_1 = 1.39\times10^{-10}/\kappa$ s. Note that according to the solution (18), in the final equilibrium state all three vectors $\boldsymbol{\alpha}$, $\boldsymbol{n}$ and $\boldsymbol{\omega}$ are arranged along the external magnetic field direction.

As we mentioned above, the numerical solution of the nonlinear set of Eqs. (13) with arbitrary initial conditions for the vectors $\boldsymbol{\alpha}$, $\boldsymbol{n}$ and $\boldsymbol{\omega}$ shows that after fast transient process a complex motion of these vectors persists that is a superposition of different oscillation modes, Eqs. (17). To study these modes individually, it is necessary to set the initial conditions for the vectors $\boldsymbol{\alpha}$, $\boldsymbol{n}$ and $\boldsymbol{\omega}$ according to Eqs. (16a). Below, we compare the approximate analytical solution, Eqs. (15) - (20), with the numerical simulation of Eqs. (13). The latter is carried out for the above values of the magnetic parameters, i.e. $K_1 = 10^5$ erg/cm$^3$, $M_s = 500$ emu/cm$^3$, $D = 5\times10^{-6}$ cm, $H_0 = 10$ Oe, and $\rho = 5$ g/cm$^3$. The magnetic damping varies within the range $\kappa = 0.01 – 1.0$. To achieve a reasonable accuracy of the numerical simulation, the numerical time step does not exceed 1/100 of the characteristic precession time of the motionless nanoparticle, i.e. $T_p \sim 1/\gamma H_k = 1.4\times10^{-10}$ s.

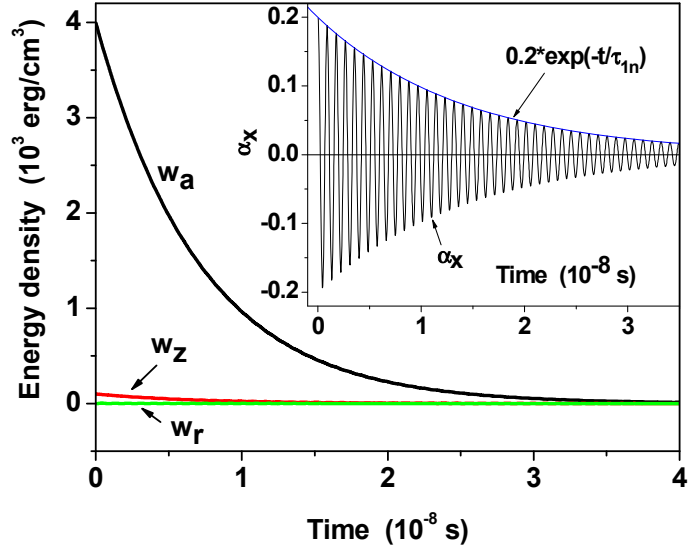

Fig. 1 The time dependence of the particular energies of magnetic nanoparticle for the high-frequency mode $\nu_1$ which corresponds to the usual ferromagnetic resonance. Inset shows the behavior of the $\alpha_x$ component of the unit magnetization vector.

Fig. 1 shows the behavior of the high-frequency mode $\nu_1$ of the magnetic nanoparticle. For the numerical simulation of this mode we choose the initial state of the vectors $\boldsymbol{\alpha}$, $\boldsymbol{n}$ and $\boldsymbol{\omega}$ at $t = 0$ setting in Eq. (15) $A_1 = 0.2$, $\nu = \nu_1$ and $\varphi = 0$, and using Eq. (16a) for the ratios $B_1/A_1$ and $C_1/A_1$. The longitudinal components of the vectors $\boldsymbol{\alpha}$ and $\boldsymbol{n}$ are determined by the normalization conditions, $\alpha_z = \sqrt{1-\alpha_x^2-\alpha_y^2}$ and $n_z = \sqrt{1-n_x^2-n_y^2}$. The $z$ component of the angular velocity of the particle is assumed to be zero, $\omega_z = 0$, so that $j_z = -\nu\alpha$. The solid curves in Fig. 1 show the time dependence of the densities of anisotropy energy, $w_a = K_1\left(1-(\boldsymbol{\alpha}\boldsymbol{n})^2\right)$, Zeeman energy, $w_z = M_s H_0 (1-\alpha_z)$, and rotational energy of the particle, $w_r = I\vec{\omega}^2/2V$, respectively. The calculations are carried out for the damping constant $\kappa = 0.01$. Inset in Fig. 1 shows the time dependence of the $x$-component of the vector $\boldsymbol{\alpha}$. As noted above, for the mode $\nu_1$ the transverse components of the vector $\boldsymbol{n}$ are negligibly small, as it is nearly parallel to the magnetic field direction.

The frequency and the relaxation time for the mode $\nu_1$ obtained numerically are in a reasonable agreement with the Eqs. (17) and (19). For example, according to Eq. (19) for $\kappa = 0.01$ and the magnetic parameters given above, the relaxation time of the mode $\nu_1$ is given by $\tau_1 = 1.39\times10^{-8}$ s, while the corresponding numerical value is $\tau_{1n} = 1.4\times10^{-8}$ s (see inset in Fig. 1). As Fig. 1 shows, for the mode $\nu_1$ the rotational energy density of the particle is negligibly small, $w_r \approx 0$, because of a small angular frequency of the particle rotation and small value of the ratio $I/V \sim 10^{-11}$ g/cm. The Zeeman energy density is also small due to small value of $H_0$. Thus, the total energy of the magnetic nanoparticle in this case is determined by the magnetic anisotropy energy.

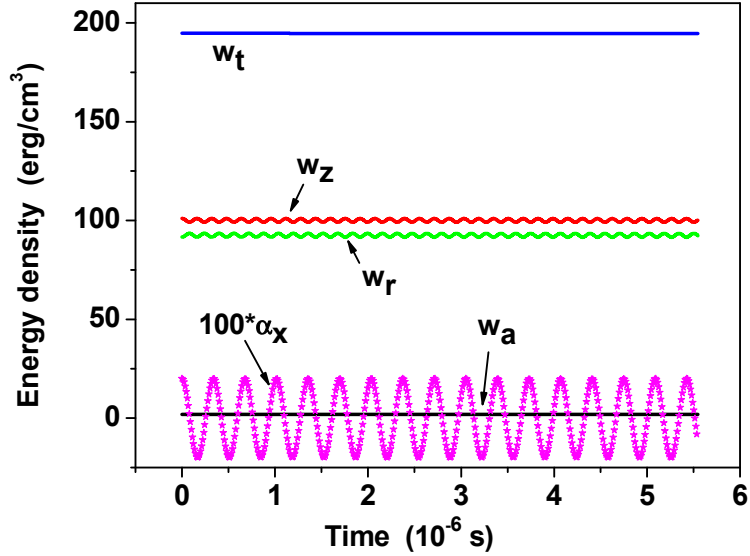

Fig. 2. The particular energies of magnetic nanoparticle for the low-frequency mode $v_2$. Note the enlarged scale of the time axis as compared with Fig. 1. The dotted line shows the $\alpha_x$ component increased by a factor 100.

In fact, the low-frequency eigen-modes $v_2$ and $v_3$ differ only by a sign of rotation of the vectors $\boldsymbol{\alpha}$ and $\boldsymbol{n}$ around the external magnetic field direction. For numerical simulation of the mode $v_2$ we choose the initial state at $t = 0$ setting in Eqs. (15), (16a) $A_2 = 0.2$, $v = v_2$ and $\varphi = 0$. Also, we set $\omega_z = 0$, as before. Fig. 2 shows the results of numerical simulation of the mode $v_2$ for the damping constant $\kappa = 0.5$ and for the same values of the other magnetic parameters. Unlike the mode $v_1$, for the mode $v_2$ the unit magnetization vector is approximately parallel to the director, $\boldsymbol{\alpha} \approx \boldsymbol{n}$. Both vectors rotate in unison around the external magnetic field direction. Therefore, as shown in Fig. 2, for the mode $v_2$ the magnetic anisotropy energy density is small, $w_a \approx 0$, whereas the Zeeman and rotational energy densities have close values, $w_r \approx w_z$. Indeed, in the linear approximation, Eqs. (15) - (17), it is easy to prove that for a given mode the ratio $w_z/w_r = (v_2/v_\xi)^2 \approx 1$. For comparison, in Fig. 2 we show also the x-component of the vector $\boldsymbol{\alpha}$ multiplied by a factor 100. As will be shown below, the small oscillations of the particular energies $w_r$ and $w_z$, visible in Fig. 2 are associated with a small contribution of the mode $v_3$ which is difficult to exclude in the numerical simulation.

As noted above, the relaxation time for the low frequency modes is very large. Indeed, using Eq. (20) one obtains $\tau_2 = 2.8 \times 10^{-3}$ s for $\kappa = 0.5$. Therefore, as Fig. 2 shows, the oscillation amplitude of the transverse components of the vectors $\boldsymbol{\alpha}$, $\boldsymbol{n}$ and $\boldsymbol{\omega}$, and all the particular energies are virtually unchanged over a large number of precession periods. As a result, it is not easy to estimate the relaxation time of the mode $v_2$ numerically.

For arbitrary initial conditions for the vectors $\boldsymbol{\alpha}$, $\boldsymbol{n}$ and $\boldsymbol{\omega}$ a complex oscillatory process occurs, which is a linear superposition of the modes (17). However, the high-frequency mode $v_1$ has a very short relaxation time. Therefore, at times $t \gg \tau_1$ a superposition of the modes $v_2$ and $v_3$ survives only. As follows from Eq. (17), these eigen frequencies can be represented as $v_2 = v - \varepsilon$ and $v_3 = -v - \varepsilon$, with $v = v_\xi/\sqrt{1+v_H/v_k}$ and $\varepsilon = 0.5 v_\alpha/(1+v_H/v_k)$, where $\varepsilon \ll v$ (we neglect the second term in $\varepsilon$ since it is significantly less then the first one). For simplicity, consider the case when the modes $v_2$ and $v_3$ have equal amplitudes, $A_2 = A_3 = A$, and the same phases, $\varphi_2 = \varphi_3 = 0$. Then, the transverse components of the unit magnetization vector are given by

$$\alpha_x = A(\cos v_2 t + \cos v_3 t) = 2A \cos vt \cos \varepsilon t ;$$
$$\alpha_y = A(\sin v_2 t + \sin v_3 t) = -2A \cos vt \sin \varepsilon t . \qquad (21a)$$

The same formulas are valid for the transverse components of the vector **n** with the replacement of $A$ for $B$. Since the coefficient $C_3 \approx -C_2$, for the transverse components of the angular velocity one obtains

$$\omega_x = 2C \sin \nu t \sin \varepsilon t; \qquad \omega_y = 2C \sin \nu t \cos \varepsilon t. \qquad (21b)$$

It follows from Eq. (21) that the transverse components of the vectors $\alpha$, **n** and $\omega$ oscillate with the frequency $\nu$ in a plane passing through the $z$-axis. Simultaneously, this plane rotates slowly around the $z$ axis with the frequency $\varepsilon$. For the densities of the Zeeman energy, the anisotropy energy and the rotation energy one obtains in this case

$$w_z = M_s H_0 (1 - \alpha_z) \approx \frac{M_s H_0}{2}(\alpha_x^2 + \alpha_y^2) = 2 M_s H_0 A^2 \cos^2 \nu t,$$
$$w_a = K_1 (1 - (\vec{\alpha}\vec{n})^2) \approx K_1 ((\alpha_x - n_x)^2 + (\alpha_y - n_y)^2) = 4 K_1 (A - B)^2 \cos^2 \nu t, \qquad (22)$$
$$w_r \approx \frac{I}{2V}(\omega_x^2 + \omega_y^2) = 2 \frac{I}{V} C^2 \sin^2 \nu t.$$

Thus, for joint excitation of the modes $\nu_2$ and $\nu_3$ the particular energies (22) oscillate with a frequency $2\nu$ around their mean values. In this process the rotational energy $w_r$ periodically transforms into the sum of the magnetic energies, $w_a + w_z$, and vice versa. However, using Eqs. (16), (17) it can be shown that neglecting the damping the total energy of the nanoparticle $w_t$ remains constant

$$w_t = w_a + w_z + w_r = 2 A^2 M_s H_0 (1 + \nu_H / \nu_k).$$

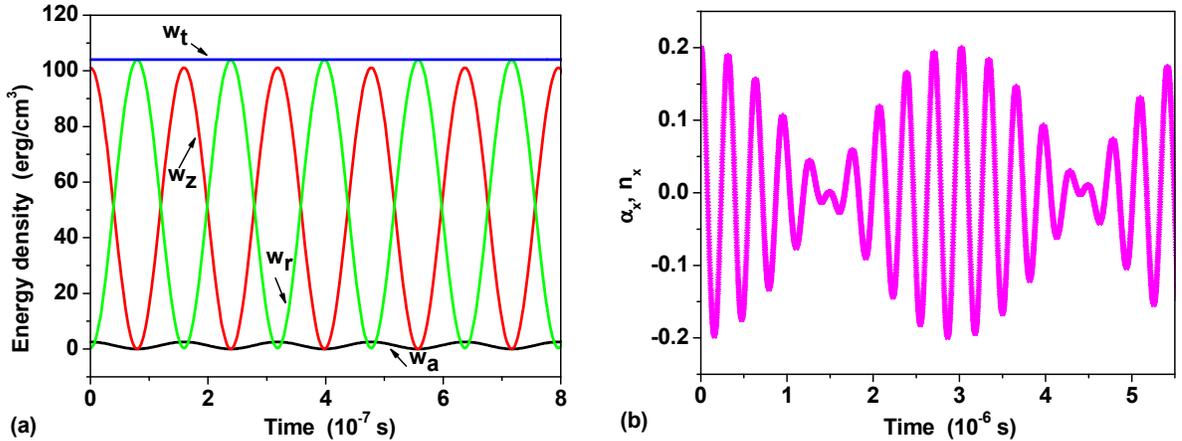

Fig. 3. The superposition of the low-frequency modes $\nu_2$ and $\nu_3$ excited with equal amplitudes and phases: a) the behavior of the particular energies, and b) the time dependence of the $x$-components of vectors $\alpha$ and **n**.

Fig. 3a shows the behavior of the particular energies of magnetic nanoparticle, and Fig. 3b shows the behavior of the unit magnetization vector and the director for the case when the modes $\nu_2$ and $\nu_3$ are excited with equal phases and amplitudes, $A_2 = A_3 = 0.1$. One can see that the numerical simulation data are in a good correspondence with Eqs. (21), (22). A similar behavior is observed also for other relations between the amplitudes and phases of the low-frequency modes $\nu_2$ and $\nu_3$.

Let us now consider briefly the magnetization process in the absence of the external magnetic field. In this case Eq. (7) leads to the conservation of the total momentum of the particle

$$\vec{J}/I = \vec{j}_0 = \vec{\omega} - \nu_\alpha \vec{\alpha} = const.$$

Consequently, the vector $\boldsymbol{\omega}$ can be excluded from Eqs. (3) and (5). This leads to the following equations for the vectors $\boldsymbol{\alpha}$ and $\boldsymbol{n}$

$$\frac{d\vec{n}}{dt} = [\vec{j}_0, \vec{n}] + v_\alpha [\vec{\alpha}, \vec{n}];\qquad(23)$$

$$\frac{d\vec{\alpha}}{dt} = -v_k(\vec{\alpha}\vec{n})[\vec{\alpha}, \vec{n}] - \kappa[\vec{\alpha}, [\vec{\alpha}, v_k(\vec{\alpha}\vec{n})\vec{n} - \vec{j}_0]].$$

As numerical simulation shows, in this case the precession of the vectors $\boldsymbol{\alpha}$ and $\boldsymbol{n}$ is around the direction of the vector $\boldsymbol{j}_0$. Setting $\boldsymbol{j}_0 = v_j \boldsymbol{k}$, one obtains from Eq. (23) the linearized equations of motion for the transverse components of the vectors $\boldsymbol{\alpha}$ and $\boldsymbol{n}$

$$\frac{dn_x}{dt} = -(v_j + v_\alpha)n_y + v_\alpha \alpha_y;\qquad \frac{dn_y}{dt} = (v_j + v_\alpha)n_x - v_\alpha \alpha_x;$$

$$\frac{d\alpha_x}{dt} = -v_k(\alpha_y - n_y) - \kappa v_k(\alpha_x - n_x) + \kappa v_j \alpha_x;\qquad(24)$$

$$\frac{d\alpha_y}{dt} = v_k(\alpha_x - n_x) - \kappa v_k(\alpha_y - n_y) + \kappa v_j \alpha_y.$$

Neglecting the magnetic damping, $\kappa \to 0$, one can easily check that the solution of Eqs. (24) has the form

$$\alpha_x = A\cos(\nu t + \varphi);\qquad \alpha_y = A\sin(\nu t + \varphi);\qquad(25)$$
$$n_x = B\cos(\nu t + \varphi);\qquad n_y = B\sin(\nu t + \varphi),$$

where the amplitudes $A$ and $B$ are related by

$$B = \frac{v_k - \nu}{v_k}A,\qquad(26a)$$

and the corresponding dispersion equation for the eigen-frequency $\nu$ reads

$$\nu^2 - \nu(v_j + v_k + v_\alpha) + v_j v_k = 0.\qquad(26b)$$

For further analysis, we assume that the characteristic frequencies $v_\alpha$ and $v_j$ are small with respect to $v_k$, i.e. $v_\alpha, v_j \ll v_k$. In this case Eq. (26b) has a high-frequency and low-frequency solutions, $\nu_1$ and $\nu_2$, respectively

$$\nu_1 = v_k\left(1 + \frac{v_\alpha}{v_k - v_j}\right);\qquad \nu_2 = v_j\left(1 - \frac{v_\alpha}{v_k - v_j}\right).\qquad(27)$$

The high-frequency oscillation mode is again close to the usual ferromagnetic resonance frequency, and for this mode the amplitude of the oscillation of the vector $\boldsymbol{n}$ is negligibly small, $B_1 \approx 0$. On the other hand for the low-frequency mode one has $B_2 \approx A_2$. Therefore the vectors $\boldsymbol{\alpha}$ and $\boldsymbol{n}$ are nearly parallel. Note also, that in the case considered the vectors $\boldsymbol{\alpha}$, $\boldsymbol{n}$ and $\boldsymbol{\omega}$ line up along the direction of the reduced total momentum $\boldsymbol{j}_0$ at the end of the relaxation process.

## IV. Magnetic particle scattering

Consider now the behavior of a magnetic nanoparticle in a non-uniform external magnetic field. In this case, the motion of the center of mass of the nanoparticle in the Cartesian coordinates $(x, y, z)$ is governed [15] by the equations

$$\rho \frac{d^2 x}{dt^2} = M_s (\vec{\alpha} \nabla) H_x; \qquad \rho \frac{d^2 y}{dt^2} = M_s (\vec{\alpha} \nabla) H_y; \qquad \rho \frac{d^2 z}{dt^2} = M_s (\vec{\alpha} \nabla) H_z. \quad (28)$$

Further we restrict ourselves to a simple case of scattering of magnetic nanoparticle on the two dimensional magnetic "lens", when the external magnetic field has the form

$$\vec{H} = (H_x(x,z), 0, H_z(x,z)). \quad (29)$$

Since the stationary magnetic field in a vacuum satisfies the Maxwell's equations, $\operatorname{div} \vec{H} = 0$, $\operatorname{rot} \vec{H} = 0$, the partial derivatives of the magnetic field components are related by

$$\frac{\partial H_x}{\partial x} = -\frac{\partial H_z}{\partial z}; \qquad \frac{\partial H_x}{\partial z} = \frac{\partial H_z}{\partial x}. \quad (30)$$

Using Eqs. (30), the $H_x$ component can be eliminated from Eqs. (28). Then one arrives to the equations

$$\frac{d^2 x}{dt^2} = \frac{M_s}{\rho}\left(-\alpha_x \frac{\partial H_z}{\partial z} + \alpha_z \frac{\partial H_z}{\partial x}\right); \qquad \frac{d^2 z}{dt^2} = \frac{M_s}{\rho}\left(\alpha_x \frac{\partial H_z}{\partial x} + \alpha_z \frac{\partial H_z}{\partial z}\right). \quad (31)$$

The motion of the particle along the $y$ axis is free. For simplicity, the $y$-component of the particle velocity may be set equal to zero.

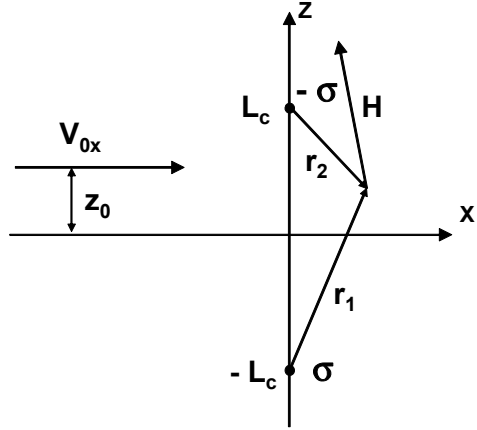

Fig. 4. The magnetic field of a two-dimensional magnetic lens.

Consider a magnetic field distribution for a specific two-dimensional lens. Let the lens consists of a line of negative magnetic poles distributed along the $y$ axis with a linear charge density $-\sigma$, crossing the plane $(x, z)$ at the point $(0, L_c)$, and of a parallel line of the positive magnetic charge of a linear density $\sigma$. The latter line intersects the plane $(x, z)$ in the symmetric point $(0, -L_c)$, as Fig.

4 shows. Then the magnetic field of the lens at an arbitrary point of the plane $(x, z)$ is determined by the equation

$$\vec{H} = 2\sigma\left(\frac{\vec{r}_1}{r_1^2} - \frac{\vec{r}_2}{r_2^2}\right), \tag{32a}$$

where the corresponding two-dimensional vectors are given by $\vec{r}_1 = (x, z + L_c)$, $\vec{r}_2 = (x, z - L_c)$. Hence the magnetic field components of the lens are

$$H_x = -H_{0z}\frac{2xzL_c^2}{(x^2 + (z+L_c)^2)(x^2 + (z-L_c)^2)};$$

$$H_z = H_{0z}\frac{L_c^2(x^2 - z^2 + L_c^2)}{(x^2 + (z+L_c)^2)(x^2 + (z-L_c)^2)}. \tag{32b}$$

Here we define the $z$-component of the magnetic field at the origin as $H_z(0,0) = 4\sigma/L_c = H_{0z}$.

When solving Eqs. (31), (32) one has to take into account that the components of the unit magnetization vector $\alpha_x$, $\alpha_z$ satisfy Eqs. (5) - (7). Thus, they change in a complex way during the translational motion of the center of inertia of the nanoparticle. The only simplification of the problem is based on the fact that due to the high precession frequency of the vector $\boldsymbol{\alpha}$ and the relatively slow motion of the nanoparticle center of mass one can assume that the precession is performed in the local magnetic field that acts on the nanoparticle at a given point of the particle trajectory.

It is useful to consider first a simpler problem of scattering on the lens (32) a "passive" magnetic moment. In this limiting case, it is assumed that the internal degrees of freedom of magnetic nanoparticle are at equilibrium, so that the vectors $\boldsymbol{\alpha}$ and $\boldsymbol{n}$ are parallel to each other and directed along the local magnetic field existing at a given point of the trajectory. In other words

$$\alpha_x = \alpha_{x,p} = H_x/\sqrt{H_x^2 + H_z^2}; \qquad \alpha_z = \alpha_{z,p} = H_z/\sqrt{H_x^2 + H_z^2}. \tag{33}$$

In this approximation the set of Eqs. (31) - (32) is closed and allows the numerical integration using, for example, a simple Euler's scheme [16].

Obviously, at fixed values of $L_c$ and $H_{0z}$, the trajectory of the passive magnetic moment is determined by the following parameters: 1) the initial velocity of the particle $V_{0x}$ along the $x$ axis, the impact parameter $z_0 < L_c$, and the ratio $M_s/\rho$. Numerical integration of Eqs. (31) - (33) shows (see Fig. 5a) that the asymptotic trajectories of the passive magnetic moment are straight lines, as it should be in the scattering.

It is easy to see that under the approximation (33) various nanoparticles differ only in the different ratios $M_s/\rho$. Therefore, it is possible to determine this ratio by measuring the scattering angle of the nanoparticle for a given initial parameters ($V_{0x}$, $z_0$). The trajectories of the passive magnetic moment are shown in Fig. 5a depending on the initial velocity $V_{0x}$ and impact parameter $z_0$ for the case of $M_s = 500$ emu/cm$^3$, $\rho = 5$ g/cm$^3$, and for the magnetic lens parameters $L_c = 1$ cm, $H_{0z} = 1000$ Oe.

Carrying out the numerical solution of Eqs. (5) - (7), (31), (32) one obtains that the actual path of the magnetic nanoparticle is very close to the trajectory of the passive moment having the same $M_s/\rho$ ratio and the same initial parameters ($V_{0x}$, $z_0$). Indeed, as Fig. 5b shows, due to the presence of the magnetic damping, and because of a relatively low particle velocity, the oscillations of the unit magnetization vector components $\alpha_x$ and $\alpha_z$ decay rapidly as a function of the $x$ coordinate. Evidently, they coincide with the corresponding values $\alpha_{x,p}$ and $\alpha_{z,p}$ of the passive moment before the particle passes the center of the magnetic lens, where the gradient of the magnetic field and the particle acceleration are maximal. The calculations in Fig. 5b are carried out for the magnetic nanoparticle with the parameters $K_1 = 10^5$ erg/cm$^3$, $M_s = 500$ emu/cm$^3$, $D = 5\times10^{-6}$ cm, $\rho = 5$ g/cm$^3$. The initial parameters are given by $V_{0x} = 500$ cm/s and $z_0 = 0.5$ cm, respectively, the damping constant is assumed to be $\kappa = 0.1$.

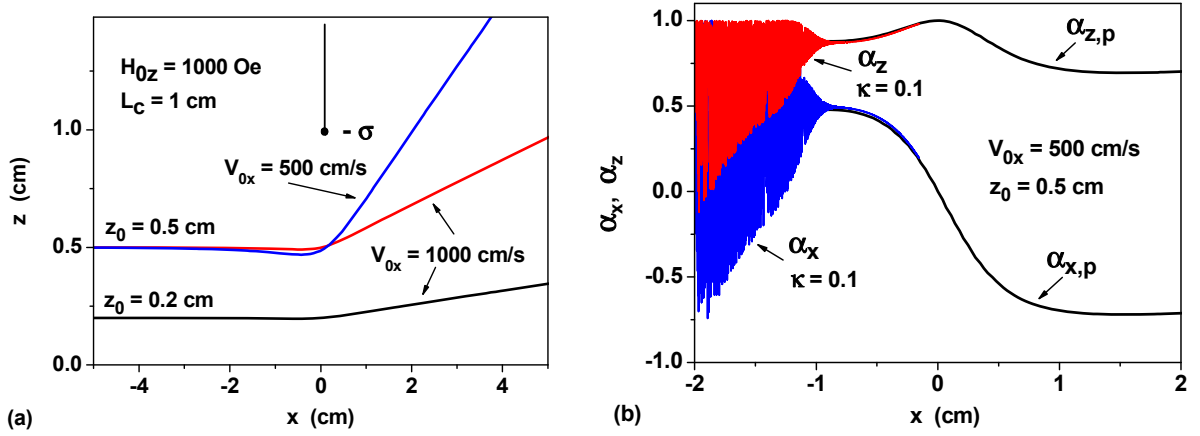

Fig. 5. a) The trajectories of a passive magnetic moment at different initial velocities $V_{0x}$ and impact parameters $z_0$; b) $\alpha_x$ and $\alpha_z$ components of the unit magnetization vector of magnetic nanoparticle as functions of the $x$ - coordinate in comparison with the corresponding components $\alpha_{x,p}$ and $\alpha_{z,p}$ of the passive magnetic moment. The initial parameters of the particle and passive moment are the same, $V_{0x} = 500$ cm/s, $z_0 = 0.5$ cm.

One can see in Fig. 5b that the behavior of the unit magnetization vector of magnetic nanoparticle differs from that of the passive moment only for the initial part of the trajectory. However, it has little influence on the scattering angle of the nanoparticle. As a result, the trajectory of magnetic nanoparticle coincides with the path of the corresponding passive moment regardless of the initial values of the vectors $\boldsymbol{\alpha}$, $\boldsymbol{n}$ and $\boldsymbol{\omega}$. The calculations show that this conclusion is also valid for sufficiently small values of the damping constant, $\kappa \sim 0.01$, because the oscillations of the unit magnetization vector of the nanoparticle are significantly suppressed when the particle passes through the center of the lens, $x \approx 0$. This is because at the lens center the large $H_z$ magnetic field component completely orients the vectors $\boldsymbol{\alpha}$ and $\boldsymbol{n}$ along the $z$ axis.

## V. Conclusions

The problem of the motion of charged particles in an external electromagnetic field is important for various areas of physics, such as solid state physics, plasma physics, astrophysics, etc. It is discussed in detail in many famous test-books [17,18] on electrodynamics. On the other hand, to the best of our knowledge, the movement of magnetic nanoparticles in an external magnetic field has not been considered so far, although this problem arises naturally in the study of the scattering of magnetic nanoparticles and magnetic nanoclusters [11-13] in non – uniform external magnetic field. In this paper we postulate the basic set of equations that describe the motion of a free magnetic nanoparticle both in uniform and non-uniform external magnetic field taking into account the conservation of the total momentum of the nanoparticle, which is the sum of the mechanical and the total spin momentum. It is shown that for a free magnetic nanoparticle there are three independent oscillation modes of the vectors $\boldsymbol{\alpha}$, $\boldsymbol{n}$ and $\boldsymbol{\omega}$, which describe the internal degrees of freedom of the nanoparticle. The high-frequency mode $\nu_1$ is similar to the conventional ferromagnetic resonance. For this mode the director $\boldsymbol{n}$ is almost parallel to the external magnetic field direction, whereas the unit magnetization vector $\boldsymbol{\alpha}$ is experiencing rapidly damped precession. However, there are also two low-frequency oscillation modes with frequencies $\nu_2$ and $\nu_3 \approx -\nu_2$. For these modes the vectors $\boldsymbol{\alpha}$ and $\boldsymbol{n}$ are nearly parallel and rotate in unison around the direction of the external magnetic field. The presence of the low-frequency oscillation modes with remarkably long relaxation time means that in a rare assembly of magnetic nanoparticles there exists the resonant absorption of the energy of alternating magnetic field at a frequency that is small compared with a conventional ferromagnetic resonance frequency. This interesting effect seems to deserve experimental confirmation.